\documentclass[conference]{IEEEtran}

\IEEEoverridecommandlockouts
\usepackage{graphicx}
\usepackage{cite}
\usepackage{amsmath}
\usepackage{amsfonts}
\usepackage{amssymb}
\usepackage{subfigure}
\usepackage{psfrag}
\usepackage{color}
\usepackage{tikz}
\usepackage{multicol}

\newcommand{\bm}{\mathbf}
\newcommand{\be}{\begin{equation}}
\newcommand{\ee}{\end{equation}}
\newcommand{\bea}{\begin{eqnarray}}
\newcommand{\eea}{\end{eqnarray}}
\newcommand{\bone}{{\bm 1}}
\newcommand{\x}{{\bm x}}

\newcommand{\y}{{\bm y}}
\newcommand{\z}{{\bm z}}

\newcommand{\br}{{\bm r}}

\newcommand{\dd}{{\bm d}}
\newcommand{\bA}{{\bm A}}

\newcommand{\bR}{{\bm R}}

\newcommand{\bF}{{\bf F}}
\newcommand{\bG}{{\bf G}}
\newcommand{\bD}{{\bf D}}

\newcommand{\bS}{{\bf S}}
\newcommand{\bH}{{\bf H}}

\newcommand{\bd}{{\bf d}}
\newcommand{\bs}{{\bf s}}

\newcommand{\bzero}{{\bf 0}}

\newcommand{\I}{{\bm I }}

\newcommand{\BU}{{\boldsymbol{\mathcal U}}}

\newcommand{\BH}{{\boldsymbol{\mathcal H}}}
\newcommand{\BD}{{\boldsymbol{\mathcal D}}}

\newcommand{\blambda}{\mbox{\boldmath$\lambda$}}
\newcommand{\bLambda}{\mbox{\boldmath$\Lambda$}}
\newcommand{\bGamma}{\mbox{\boldmath$\Gamma$}}
\newcommand{\bOmega}{\mbox{\boldmath$\Omega$}}
\newcommand{\bPsi}{\mbox{\boldmath$\Psi$}}

\newcommand{\bpsi}{\mbox{\boldmath$\psi$}}

\newcommand{\bomega}{\mbox{\boldmath$\omega$}}

\title{SC-FDMA as a Delay-Doppler Domain \\ Modulation Technique}

\author{
\IEEEauthorblockN{Arman Farhang and Mohsen Bayat} 
\IEEEauthorblockA{Department of Electronic \& Electrical Engineering, Trinity College Dublin, Ireland \\
\{arman.farhang,bayatm\}@tcd.ie}
\vspace{-1cm}

}

\begin{document}
\maketitle

\begin{abstract}
This paper compares orthogonal time frequency space (OTFS) modulation and single-carrier frequency division multiple access (SC-FDMA). It shows that these are equivalent except for a set of linear phase shifts, applied to the transmit/receive data symbols, which can be absorbed into the channel. Through mathematical and numerical analysis, it is confirmed that SC-FDMA is in fact a delay-Doppler domain multiplexing technique that can achieve the same performance gains as those of OTFS in time-varying wireless environments. 
This is a promising result as SC-FDMA is already a part of the current wireless standards. 
The derivations in this paper also shed light on the time-frequency resources used by the delay-Doppler domain data symbols with the fine granularity of delay and Doppler spacings. While comparing the detection performance of the two waveforms, a timing offset (TO) estimation technique with orders of magnitude higher accuracy than the existing solutions in the literature is proposed. From multiple access viewpoint, the underlying tile structures in the time-frequency domain for OTFS and SC-FDMA are discussed. Finally, multiuser input-output relationships for both waveforms in the uplink are derived.

\end{abstract}

\vspace{-2 mm}
\section{Introduction}\label{sec:Introduction}
One of the main challenges in the next generation wireless networks is provision of reliable communications to high mobility users \cite{Wei2021}. The high mobility and high data rate requirements lead to highly selective wireless channels in time and frequency, up to a level that orthogonal frequency division multiplexing (OFDM) cannot handle \cite{Wei2021}. To combat the detrimental effects of such channels, orthogonal time frequency space (OTFS) modulation \cite{OTFS}, emerged as a paradigm-shifting technology. OTFS 
has opened less explored avenues in wireless communications through delay-Doppler signaling. 

In its initial proposal, OTFS was presented as a 2D precoded OFDM system \cite{OTFS}. This implies a close relationship between this waveform and discrete Fourier transform (DFT) precoded OFDM, also known as single-carrier frequency division multiple access (SC-FDMA) \cite{Goodman2006}. Using this relationship, the authors in \cite{Andrew2021_2} studied OTFS in the frequency-Doppler domain and developed a channel estimation method by utilizing a frequency-domain pilot. While the authors in \cite{Andrew2021_2} implicitly touched upon the similarities of OTFS and DFT-precoded OFDM, they did not investigate whether the same gains as those of OTFS can be achieved by DFT-precoded OFDM. 

The main thrust of this paper is to study and compare OTFS and SC-FDMA both through mathematical derivations and simulations to find out whether either of them has any advantages over the other one. 
An interesting outcome of our study is that essentially these waveforms are equivalent as they only differ in a set of phase-shifts, applied to the transmit and receive data symbols, which can be absorbed into the channel. 
Hence, SC-FDMA can be seen as a delay-Doppler multiplexing technique, if it is looked at from the perspective of the DFT-precoders inputs and the inverse DFT (IDFT) post-processing units outputs. 
This is while SC-FDMA literature is mostly focused on frequency domain training, processing and equalization \cite{Falconer2002,Liu2015,Kiayani2016}.
Through simulations, we confirm that SC-FDMA achieves the same performance as that of OTFS, by using the delay-Doppler domain pilots \cite{Raviteja2019}, synchronization \cite{Bayat2022}, channel estimation \cite{Raviteja2019}, and equalization techniques \cite{LSMR2021}.
Our derivations shed light on the time-frequency resources used by the delay-Doppler domain data symbols with the fine granularity of delay and Doppler spacings. This is quite important from the backward compatibility, coexistence and resource allocation viewpoints. As another contribution of this paper, we propose a timing offset (TO) estimation technique with orders of magnitude higher accuracy than the solution in \cite{Bayat2022}. Finally, we discuss multiple access aspects of the waveforms under study while providing input-output relationships for the uplink communication with generalized resource allocation.

\textit{Notations}:
Scalar values, vectors, and matrices are denoted by normal, boldface lowercase, and boldface uppercase letters, respectively. 
The function ${\rm{diag}} \{ \x \}$ forms a diagonal matrix with the elements of $\x$ on its main diagonal.
$\bm{1}_{N}$ and $\bm{0}_{N}$ are all-ones and zero vectors of length $N$, respectively.
$\bm{I}_{N}$ is the identity matrix of size $N$ and $\bm{0}_{M\times N}$ is an $M\times N$ zero matrix.  
The superscripts $(.)^{\rm{H}}$ and $(.)^{\rm{T}}$ indicate Hermitian and transpose operations, respectively, and
$\otimes$ denotes Kronecker product.
$\bm{F}_N$ is the normalized $N$-point DFT matrix with the elements $[\bm{F}_N]_{p,q}=\frac{1}{\sqrt{N}} e^{-j \frac{2 \pi pq}{N}}$ for $p,q=0,\ldots, N-1$.

\section{OTFS Modulation}\label{sec:OTFS_Mod}
\begin{figure}
\psfrag{x}{\small $\x$}
\psfrag{d0}{\scriptsize $\bd_0$}
\psfrag{dM-1}{\scriptsize $\bd_{M-1}$}
\psfrag{s0}{\scriptsize $\bs_0$}
\psfrag{sM-1}{\scriptsize $\bs_{M-1}$}
\psfrag{d[n]}{\hspace{-1mm}\small$d[n]$}
\psfrag{M}{\small $M$}
\psfrag{s}{\scriptsize $\bs$}
\psfrag{F_NH}{\small $\bF_N^{\rm H}$}
\psfrag{S/P}{\hspace{-1.3 mm}S/P}
\psfrag{Circ}{\small CircShift}
\psfrag{M-1}{{\hspace{-0.6 mm}\small $M-1$} }
\psfrag{0}{{\hspace{-0.6 mm}\small $0$} }
\psfrag{CP}{\small CP}
\psfrag{Add}{\small Addition}
\centering
\includegraphics[scale=0.29]{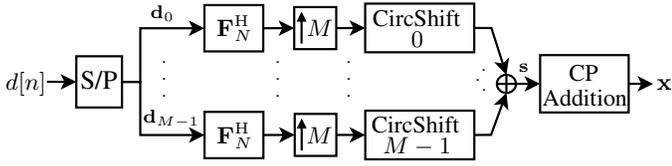} 
\vspace{-2 mm}
\caption{OTFS modulator structure.}
\vspace{-2 mm}
\label{fig:OTFS_Tx}
\end{figure}

Consider transmit quadrature amplitude modulated (QAM) data symbols taken from a zero mean independent and identically distributed (i.i.d) process with unit variance. 
In OTFS, the transmit data symbols are placed on a uniform grid in the delay-Doppler domain with delay and Doppler spacings of $\Delta\tau$ and $\Delta\nu$, respectively. 
Assuming $M$ delay bins and $N$ Doppler bins on the grid, the data symbols $d_{m}[n]$ for $m\!=\!0,\ldots,M\!-\!1$ and $N\!=\!0,\ldots,N\!-\!1$ are first transformed to the delay-time domain to form the OTFS transmit signal. This involves $N$-point IDFT operations on the rows of the data matrix $\bD\!=\![\bd_0,\ldots,\bd_{M-1}]^{\rm T}$, where $\bd_m\!=\![d_{m}[0],\ldots,d_{m}[N-1]]^{\rm T}$, i.e., 
\be\label{eqn:S}
\bS=\bD\bF_N^{\rm H}.
\ee
Then, the columns of $\bS$ are concatenated and the resulting signal is stored as a vector $\bs={\rm vec}\{\bS\}$. 
Let us consider $\bS=[\bs_0,\ldots,\bs_{M-1}]^{\rm T}$ with the rows $\bs_m^{\rm T}=[s_m[0],\ldots,s_m[N-1]]$ for $m=0,\ldots,M-1$, where $\bs_m=\bF_N^{\rm H}\bd_m$.
Concatenating the columns of $\bS$ is equivalent to interleaving its rows to form the $MN\times 1$ vector $\bs$.
In other words, $\bs$ can be formed by adding $M$-fold expanded versions of the rows in $\bS$, i.e., $\bs_m^{\rm e}=[s_m[0],\bzero_{M-1}^{\rm T},s_m[1],\bzero_{M-1}^{\rm T},\ldots,s_m[N-1],\bzero_{M-1}^{\rm T}]^{\rm T}$, that are circularly sifted by $m$ samples. This process is shown in Fig.~\ref{fig:OTFS_Tx} and can be mathematically formulated as
\be\label{eqn:s}
\bs = \sum_{m=0}^{M-1}{\rm{CircShift}}(\bs_m^{\rm e},m).
\ee
Finally, a CP with the length of $L_{\rm cp}$ that is longer than the channel delay spread is appended at the beginning of each OTFS block to avoid inter-block interference \cite{PracticalPulse}. Hence, the OTFS transmit signal can be represented as 
\be\label{eqn:x}
\x=\bA_{\rm cp}\bs,
\ee
where $\bA_{\rm cp}=[\bG_{\rm cp}^{\rm T},\I_{MN}^{\rm T}]^{\rm T}$ is the CP addition matrix and the $L_{\rm cp}\times MN$ matrix $\bG_{\rm cp}$ is comprised of the last $L_{\rm cp}$ rows of the identity matrix $\I_{MN}$.

\begin{figure}
\psfrag{dhat0}{\scriptsize $\widetilde{\bd}_0$}
\psfrag{dhatM-1}{\scriptsize $\widetilde{\bd}_{M-1}$}
\psfrag{r0}{\hspace{1.5mm}\scriptsize {$\br_0$}}
\psfrag{rM-1}{\scriptsize {$\br_{M-1}$}}
\psfrag{s0}{\scriptsize $\bs_0$}
\psfrag{sM-1}{\scriptsize $\bs_{M-1}$}
\psfrag{r[k]}{\hspace{-1mm}\small$r[\kappa]$}
\psfrag{M}{\small $M$}
\psfrag{s}{\scriptsize $\bs$}
\psfrag{F_NH}{\small $\bF_N$}
\psfrag{Circ}{\hspace{0.6 mm}\small CircShift}
\psfrag{-M+1}{{\hspace{-2 mm}\small $-(M-1)$} }
\psfrag{0}{{\hspace{0.1 mm}\small $0$} }
\psfrag{CP}{\small CP}
\psfrag{Remove}{\hspace{0.2 mm}\small Remove}
\centering
\includegraphics[scale=0.29]{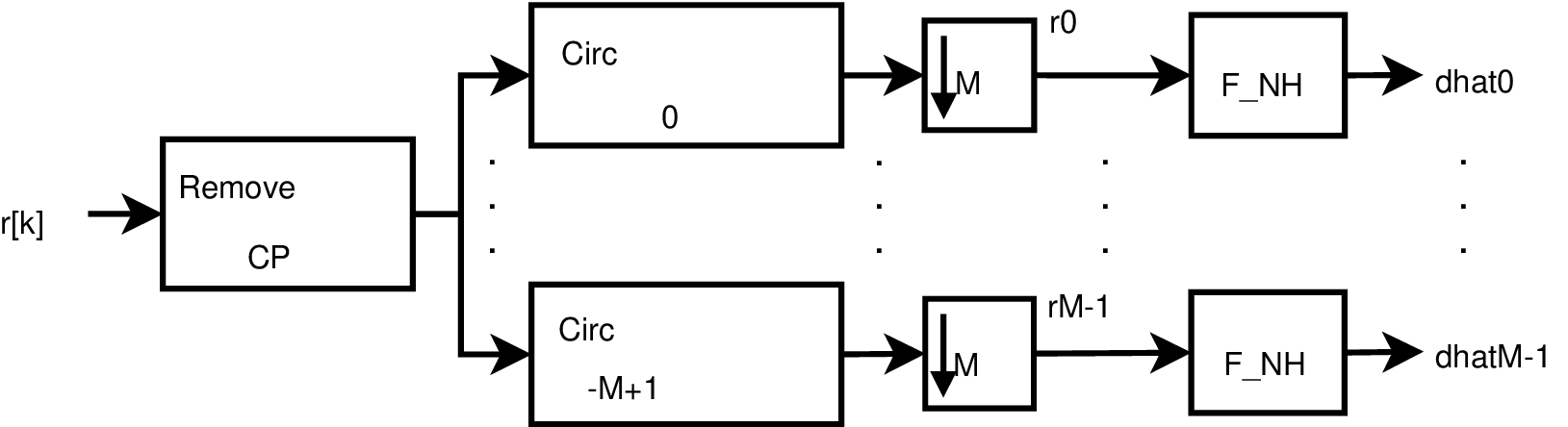} 
\vspace{-2 mm}
\caption{OTFS demodulator structure.}
\vspace{-5 mm}
\label{fig:OTFS_Rx}
\end{figure}

After the transmit signal $x[\kappa]$ is passed through the linear time varying (LTV) channel with length $L_{\rm ch}$, assuming perfect synchronization, the received signal can be obtained as
\be\label{eqn:r}
r[\kappa]=\sum_{\ell=0}^{L_{\rm ch}-1}h[\ell,\kappa]x[\kappa-\ell]+\eta[\kappa],
\ee
where $h[\ell,\kappa]$ is the channel gain at the delay tap $\ell$ and sample $\kappa$, and $\eta[\kappa]\sim\mathcal{CN}(0,\sigma_\eta^2)$ is complex normal additive white Gaussian noise (AWGN) with the variance $\sigma_\eta^2$. After discarding the CP, the received OTFS signal in the delay-time domain can be formed into an $M\times N$ matrix $\bR=[\br_0,\ldots,\br_{M-1}]^{\rm T}$ where $\br_m=[r[L_{\rm cp}+m],r[L_{\rm cp}+m+M],\ldots,r[L_{\rm cp}+m+(N-1)M]]^{\rm T}$.
As shown in \cite{farhang2017low}, the received signal in the delay-Doppler domain can be obtained from the delay-time domain signal by taking DFT from the rows of $\bR$, i.e.,
\be\label{eqn:}
\widetilde{\bD}=\bR\bF_{N},
\ee
where $\widetilde{\bD}=[\widetilde{\dd}_0,\ldots,\widetilde{\dd}_{M-1}]^{\rm T}$ and $\widetilde{\dd}_m=\bF_N \br_m$ with the elements $\widetilde{d}_m[n]$ for $n=0,\ldots,N-1$. Choosing the samples across the $m^{\rm th}$ row of $\bR$, i.e., $\br_m^{\rm T}$, is equivalent to circularly shifting the received OTFS signal samples by $m$ positions after CP removal and down-sampling by a factor of $M$.  
Hence, OTFS demodulator can be implemented using the structure that is shown in Fig.~\ref{fig:OTFS_Rx}.

\section{Relationship Between OTFS and SC-FDMA}
\label{sec:OTFS_SC-FDMA} 
In this section, we present derivations that reveal the relationship between OTFS and SC-FDMA. This leads to deep insights into the time-frequency resources that are occupied by each delay-Doppler data symbol that are important from resource allocation viewpoint.  

\begin{figure}
\psfrag{F_NH}{ \hspace{-0.7mm}$\bF_N^{\rm H}$}
\psfrag{F_MNH}{ \hspace{-2.3 mm}$\bF_{MN}^{\rm H}$}
\psfrag{M}{{$M$} }
\psfrag{dm}{\small{$\bd_m$} }
\psfrag{dmp}{\small{$\bd_m^{\rm p}$} }
\psfrag{s_e}{\small{$\bs_m^{\rm e}$} }
\centering
\includegraphics[scale=0.35]{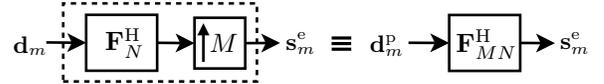} 
\vspace{-2 mm}
\caption{Upsampling in the frequency vs. time domain.}
\vspace{-2 mm}
\label{fig:Upsampling}
\end{figure}

Firstly, let us take a closer look at a given branch, $m$, of the OTFS modulator in Fig.~\ref{fig:OTFS_Tx}. 
From multirate signal processing theory \cite{Vaidyanathan1993}, $M$-fold upsampling of the signal $\bs_m$ is equivalent to creating $M$ spectral replicas of the signal in the frequency domain, i.e., $M$ repetitions of the Doppler domain data symbols in each delay bin. 
Hence, the $N$-point IDFT and $M$-fold upsampling operations on the branch $m$ of the OTFS modulator is equivalent to taking $MN$-point IDFT from the periodic signal $\bd_m^{\rm p}\!=\!\frac{1}{\sqrt{M}}(\bone_M\otimes \bd_m)$, see Fig.~\ref{fig:Upsampling}. 

Moreover, due to the circular shift property of the DFT, circularly shifting the sequence $\bs_m^{\rm e}$ by $m$ samples in Fig.~\ref{fig:OTFS_Tx} is equivalent to element-wise multiplication of the vector $\bd_m^{\rm p}$ by $\blambda_m=[1,e^{-j\frac{2\pi m}{MN}},\ldots,e^{-j\frac{2\pi m}{MN}(MN-1)}]^{\rm T}$.
Consequently, (\ref{eqn:s}) can be rearranged as
\be\label{eqn:s_dft}
\bs
=\sum_{m=0}^{M-1}\bF_{MN}^{\rm H}\bLambda_m\bd_m^{\rm p}=\bF_{MN}^{\rm H}\overline{\bd},
\ee
where $\bLambda_m={\rm diag}\{\blambda_m\}$ and $\overline{\bd}=\sum_{m=0}^{M-1}\bLambda_m\bd_m^{\rm p}$. Substituting (\ref{eqn:s_dft}) into (\ref{eqn:x}), the OTFS transmit signal can be formed as $\x=\bA_{\rm cp}\bF_{MN}^{\rm H}\overline{\bd}$. This formulation is the same as that of an OFDM modulator with $MN$ subcarriers where the QAM data symbols are spread across the frequency domain. Therefore, in the following, we focus on the details of this spreading that reveal the relationship between OTFS and DFT-precoded OFDM/SC-FDMA.

\begin{figure}
\psfrag{F_M}{ \hspace{-1.4mm}$\bF_M$}
\psfrag{F_MN}{ \hspace{-1.7 mm}$\bF_{MN}^{\rm H}$}
\psfrag{M}{{$M$} }
\psfrag{x}{{\small$\x$} }
\psfrag{d0}{\hspace{-0.2mm}\scriptsize{$d_0[0]$} }
\psfrag{d1}{\hspace{-0.4mm}\scriptsize{$d_{M\!-\!1}[0]$} }
\psfrag{d4}{\hspace{-2.2mm}\scriptsize{$d_0[N\!-\!1]$} }
\psfrag{d5}{\hspace{-5mm}\scriptsize{$d_{M\!-\!1}[N\!-\!1]$} }
\psfrag{e0}{\hspace{-0.3  mm}\scriptsize{$\omega_0^0$} }
\psfrag{e1}{\hspace{-0.7 mm}{\scriptsize$\omega_0^{\hspace{-0.3mm}M\!-\!1}$} }
\psfrag{e2}{\hspace{-0.3  mm}\scriptsize{$\omega_{1}^0$} }
\psfrag{e3}{\hspace{-0.7  mm}{\scriptsize$\omega_{1}^{M-1}$} }
\psfrag{e4}{\hspace{-0.7 mm}\scriptsize{$\omega_{\hspace{-0.3mm}N\!-\!1}^0$} }
\psfrag{e5}{\hspace{-1.2 mm}{\scriptsize$\omega_{\hspace{-0.3mm}N\!-\!1}^{\hspace{-0.3mm}M\!-\!1}$} }
\psfrag{p/s}{\footnotesize{P/S} }
\psfrag{and}{\footnotesize{\&} }
\psfrag{CP}{\hspace{-0.4  mm}\footnotesize{CP} }
\psfrag{Add}{\hspace{-0.3mm}\footnotesize{Addition} }
\centering
\includegraphics[scale=0.275]{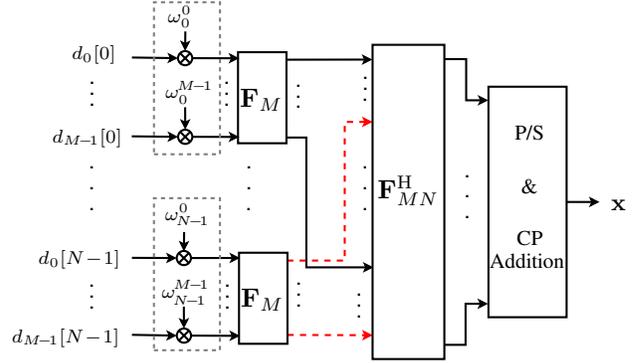} 
\vspace{-2 mm}
\caption{An alternative structure for OTFS modulator.}
\vspace{-5 mm}
\label{fig:OTFS_Tx1}
\end{figure}
Due to the periodic structure of the vectors $\bd_m^{\rm p}$, their elements are represented as $d_m^{\rm p}[n+m'N]=\frac{1}{\sqrt{M}}d_m[n],~\forall m'=0,\ldots,M-1$ and $n=0,\ldots,N-1$. Consequently, the elements of the vector $\overline{\bd}$ in equation (\ref{eqn:s_dft}) can be obtained as 
\bea\label{eqn:dbar}
\!\!\bar{d}[n+m'N]\!\!\!\!&=&\!\!\!\!\!\!\sum_{m=0}^{M-1}\!\!d_m^{\rm p}[m'N+n]e^{-j\frac{2\pi}{MN}m(n+m'N)}\nonumber\\
&=&\!\!\!\!\frac{1}{\sqrt{M}}\!\!\sum_{m=0}^{M-1}\!\!\breve{d}_m[n]e^{-j\frac{2\pi}{M}mm'},
\eea
for $n=0,\ldots,N-1$ and $m'=0,\ldots,M-1$, where $\breve{d}_m[n]=d_m[n]\omega_n^m$ and $\omega_n^m=e^{-j\frac{2\pi}{MN}mn}$. Thus, $\overline{\bd}$ is formed by taking $M$-point DFT from the modulated data symbols in each column, $n$, of $\bD$ by the carrier $\omega_n^m$ and interleaving the resulting signal samples.  

Based on (\ref{eqn:s_dft}) and (\ref{eqn:dbar}), the OTFS modulator can be implemented as shown in Fig.~\ref{fig:OTFS_Tx1}, as an alternative to the one in Fig.~\ref{fig:OTFS_Tx}. Using this structure, the delay domain data symbols in a given Doppler bin, $n$, are first modulated to the corresponding Doppler frequency, $\frac{2\pi}{MN}n$. Then, they are transformed to the frequency domain and placed in the corresponding equally spaced frequency bins that are $N$ Doppler spacings away from one another, i.e., $\Delta f \!=\!N\Delta\nu\!=\!\frac{1}{T}$ where $T\!=\!M\Delta\tau$. This leads to the formation of the frequency-Doppler domain signal, $\overline{\bd}$, which is converted to the delay-time domain by an $MN$-point IDFT operation. Finally, a CP is appended to the delay-time domain signal to obtain the OTFS transmit signal.

\begin{figure}
\psfrag{f}{ \Large\hspace{-1.8mm}$f$}
\psfrag{v}{ \Large\hspace{-1.8mm}$\nu$}
\psfrag{t}{ \Large\hspace{-1.8mm}$t$}
\psfrag{s}{ \tiny\hspace{-1.3mm}$\Delta\tau$}
\psfrag{n}{ \tiny\hspace{-1.3mm}$\Delta\nu$}
\psfrag{T}{ \tiny\hspace{-1.3mm}$T\!\!=\!\!M\!\Delta\tau$}
\psfrag{V}{\small \hspace{-0 mm}$T$}
\psfrag{z}{\small \hspace{-1.3mm}$NT$}
\psfrag{W}{ \hspace{-1.3mm}$\tau$}
\psfrag{G}{ \hspace{-1.3mm}$\nu$}
\psfrag{y}{ \hspace{-1.3mm}$f$}
\psfrag{C}{ \hspace{-1.3mm}$t$}
\psfrag{O}{\small \hspace{-2mm}$\Delta f$}
\psfrag{B}{ \small\hspace{-1.3mm}$M\Delta f$}
\psfrag{D}{ \scriptsize\hspace{-1.3mm}$\Delta f\!=\!N\Delta\nu$}
\psfrag{E}{ \scriptsize\hspace{-1.45mm}$=\!\frac{1}{T}$}
  \centering 
  \includegraphics[scale=0.33]{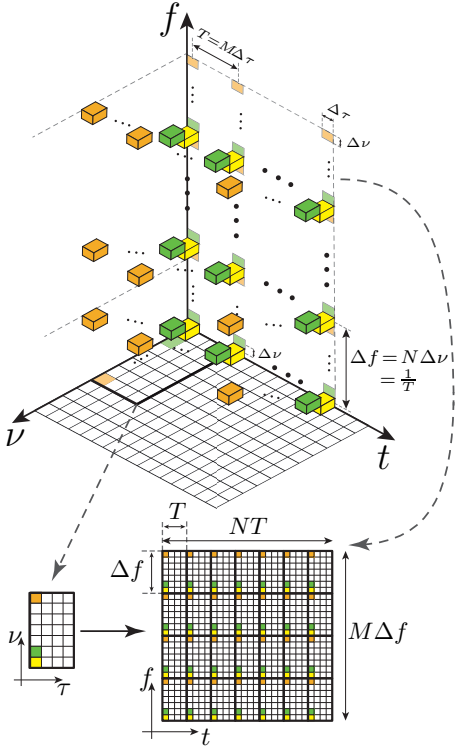}
  \vspace{-4 mm}
  \caption{Time-frequency resources used by the delay-Doppler data symbols.}
  \vspace{-5 mm}
  \label{fig:DD_TF}
\end{figure}
An important aspect of OTFS which is not very well highlighted in the literature is the time-frequency resources that are occupied by each delay-Doppler domain data symbol. Using the modulator structures in Figs.~\ref{fig:OTFS_Tx} and \ref{fig:OTFS_Tx1}, this point can be made clear. Let us consider only one active delay-Doppler domain data symbol, $d_0[0]$.
Using the structure in Fig.~\ref{fig:OTFS_Tx}, this data symbol is directly translated to the delay-time domain and it occupies $N$ equally spaced delay-time domain samples, in delay bin $0$, that are $M$ samples away from each other.
The structure in Fig.~\ref{fig:OTFS_Tx1} shows how this data symbol is spread in the frequency domain, i.e., in $M$ equally spaced frequency bins with the spacing $\Delta f$.
Due to the superposition principle and orthogonality of the delay-Doppler resources, as shown in Fig.~\ref{fig:DD_TF}, the same theory is valid for all the delay-Doppler resources.
This is extremely important from the backward compatibility, coexistence, and multiple access viewpoints as will be discussed in Section~\ref{sec:Multi_User_v0}.

\begin{figure}
\psfrag{F_NH}{ \hspace{-0.7mm}$\bF_N$}
\psfrag{F_NMH}{ \hspace{-2.3 mm}$\bF_{MN}$}
\psfrag{M}{{$M$} }
\psfrag{zm}{\small{$\z_m$} }
\psfrag{zm_f}{\small\hspace{0.3 mm}{$\bar{\z}_m$} }
\psfrag{dt_m}{\small{$\widetilde{\bd}_m$} }
\psfrag{Alias_M}{\hspace{-1mm}\scriptsize{${\rm Alias}_M(\cdot)$} }
\centering
\includegraphics[scale=0.365]{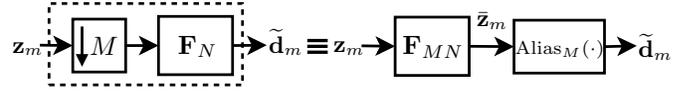} 
\caption{Downsampling in the frequency vs. time domain.}
\vspace{-3 mm}
\label{fig:Downsampling}
\end{figure}

To derive an alternative demodulator structure to the one in Fig.~\ref{fig:OTFS_Rx}, we start from the spectral effects of downsampling operation. According to the aliasing theorem, downsampling in time domain is equivalent to aliasing in the frequency domain \cite{Vaidyanathan1993}. Thus, $M$-fold downsampling and $N$-point DFT blocks in the branchs of Fig.~\ref{fig:OTFS_Rx} can be replaced by $MN$-point DFT and aliasing operations, respectively. This process is illustrated in Fig.~\ref{fig:Downsampling} where the input signals to the downsampling blocks, $m$, are represented as $\z_m={\rm CircShift}(\z,-m)$, $\z={\rm vec}\{\bR\}$ is the received signal after CP removal and $\bar{\z}_m=\bF_{MN}\z_m$. 
${\rm Alias}_{M}(\cdot)$ is the aliasing operator that performs aliasing on the length $MN$ signal $\bar{\z}_m=[\bar{z}_m[0],\ldots,\bar{z}_m[MN-1]]^{\rm T}$. This operation results in a length $N$ signal $\widetilde{\bd}_m={\rm Alias}_{M}(\bar{\z}_m)$ with the elements 
\bea\label{eqn:aliasing}
\widetilde{d}_m[n]\!\!\!\!&=&\!\!\!\!\!\frac{1}{\sqrt{M}}\sum_{m'=0}^{M-1}\bar{z}_m[n+m'N]\nonumber\\
&=&\!\!\!\!\!\left(\frac{1}{\sqrt{M}}\sum_{m'=0}^{M-1}\bar{z}[n+m'N]e^{j\frac{2\pi}{M}mm'}\right)e^{j\frac{2\pi}{MN}mn},
\eea
for $n\!=\!0,\ldots,N-1$ and $m\!=\!0,\ldots,M-1$. Due to the shift property of DFT, $\bar{\z}_m=\bLambda_m^*\bar{\z}$ where $\bar{\z}=\bF_{MN}\z$. Therefore, $\bar{z}_m[n+m'N]=\bar{z}[n+m'N]e^{j\frac{2\pi}{MN}m(n+m'N)}$. 

\begin{figure}
\psfrag{F_MH}{ \hspace{-1.3mm}$\bF_M^{\rm H}$}
\psfrag{F_MN}{ \hspace{-2.7 mm}$\bF_{MN}$}
\psfrag{M}{{$M$} }
\psfrag{r}{{\small$\br$} }
\psfrag{d0}{\hspace{-0.5  mm}\scriptsize{$\widetilde{d}_0[0]$} }
\psfrag{d1}{\hspace{-0.5  mm}\scriptsize{$\widetilde{d}_{M-1}[0]$} }
\psfrag{d2}{\hspace{-0.5  mm}\scriptsize{$\widetilde{d}_0[1]$} }
\psfrag{d3}{\hspace{-0.5  mm}\scriptsize{$\widetilde{d}_{M-1}[1]$} }
\psfrag{d4}{\hspace{-0.5  mm}\scriptsize{$\widetilde{d}_0[N-1]$} }
\psfrag{d5}{\hspace{-0.5  mm}\scriptsize{$\widetilde{d}_{M-1}[N-1]$} }
\psfrag{e0}{\hspace{-0.3  mm}\scriptsize{$\large(\hspace{-0.2  mm}\omega_{0}^{0}\hspace{-0.2  mm}\large)^*$} }
\psfrag{e1}{\hspace{-0.4 mm}{\scriptsize$\large(\hspace{-0.4  mm}\omega_{0}^{\!M-1}\hspace{-0.4  mm}\large)^*$} }
\psfrag{e2}{\hspace{-0.3  mm}\scriptsize{$\large(\hspace{-0.2  mm}\omega_{1}^{0}\hspace{-0.2  mm}\large)^*$} }
\psfrag{e3}{\hspace{-0.7  mm}{\scriptsize$\large(\hspace{-0.2  mm}\omega_{1}^{\!M-1}\!\large)^*$} }
\psfrag{e4}{\hspace{-0.7 mm}\scriptsize{$\large(\hspace{-0.2  mm}\omega_{\!N-1}^{0}\hspace{-0.2  mm}\large)^*$} }
\psfrag{e5}{\hspace{-0.7 mm}{\scriptsize$\large(\hspace{-0.4  mm}\omega_{\!N-1}^{\!M-1}\hspace{-0.4  mm}\large)^*$} }
\psfrag{p/s}{\hspace{-0.4  mm}\footnotesize{S/P} }
\psfrag{and}{\hspace{0.4  mm}\footnotesize{\&} }
\psfrag{CP}{\hspace{-1.3  mm}\footnotesize{CP} }
\psfrag{Rmv}{\hspace{-1.2  mm}\footnotesize{Remove} }
\centering
{\hspace{-9 mm}\includegraphics[scale=0.275]{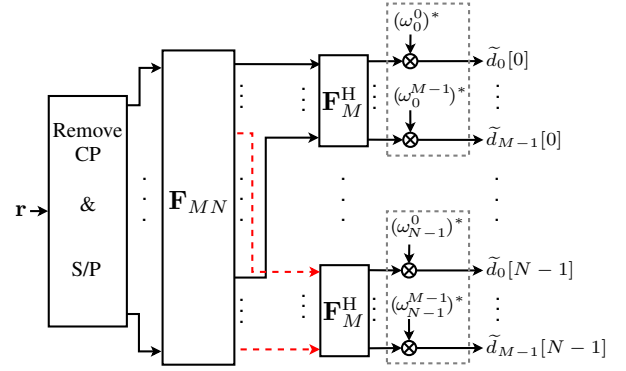} }
\caption{An alternative structure for OTFS demodulator.}
\vspace{-5 mm}
\label{fig:OTFS_Rx1}
\end{figure}
From (\ref{eqn:aliasing}), the received signals at a given Doppler bin $n$ and all the delay bins can be obtained by taking $M$-point IDFT from equally spaced elements of $\bar{\z}$, i.e., $\bar{z}[n+m'N]$ for $m'=0,\ldots,M-1$, that are demodulated by the carrier $(\omega_n^m)^*$. Thus, OTFS demodulator can be implemented as shown in Fig.~\ref{fig:OTFS_Rx1}, as an alternative to the one in Fig.~\ref{fig:OTFS_Rx}.

The structures in Figs.~\ref{fig:OTFS_Tx1} and \ref{fig:OTFS_Rx1} show the close relationship between OTFS and SC-FDMA with interleaved subcarrier allocation, known as SC-IFDMA. These architectures reveal that the only difference between OTFS and SC-IFDMA lies in the application of linear phase shifts at the input of the DFT precoders and the output of the IDFT post-processing units that are shown by dashed boxes in Figs.~\ref{fig:OTFS_Tx1} and \ref{fig:OTFS_Rx1}, respectively. The main question that remains here is \textit{if application of these phase shifts lead to any advantages for OTFS over SC-IFDMA}. Hence, to answer this question, in the subsequent sections, we compare the end-to-end channel effect and the detection performance of both waveforms.

\section{Channel Effect}
\label{sec:Channel_Effect}
The equivalent channel in the delay-Doppler domain for OTFS can be obtained through the following steps. Substituting (\ref{eqn:s_dft}) in (\ref{eqn:x}) and using (\ref{eqn:r}), the received signal after CP removal can be obtained as
\be \label{eqn:dt_rec1}
\bm{r} = \BH \bm{F}^{\rm{H}}_{MN} \bPsi (\bm{I}_{N} \otimes \bm{F}_{M}) \bOmega \bm{d} + \boldsymbol{\eta},
\ee
where $\boldsymbol{\eta} =[\eta[L_{\rm cp}],\ldots,\eta[L_{\rm cp}+MN-1]]^{\rm T}$, $\dd\!\!=\!\!{\rm vec}\{\bD\}$, $\bOmega\!\!=\!\!{\rm{diag}} \{\bomega_0,\ldots,\bomega_{N-1}\}$ and $\bomega_n\!\!=\!\![\omega_n^0,\ldots,\omega_n^{M-1}]^{\rm T}$. $\bPsi=[\bPsi_0,\ldots,\bPsi_{M-1}]^{\rm{T}}$ is the interleaving matrix in which $\bPsi_m= {\rm{CircShift}} \big( (\bm{I}_N \otimes \boldsymbol{\psi}),m \big)$ for $m=0,\ldots,M-1$ and $\bpsi=[1,\bm{0}_{M-1}]$. The channel matrix including the CP addition and removal effects is represented as $\BH = \bm{R}_{\rm{cp}} \bm{H} \bm{A}_{\rm{cp}}$ where $\bm{R}_{\rm{cp}}=[\bm{0}_{MN \times L_{\rm{cp}}},\bm{I}_{MN}]$ is the CP removal matrix and $\bm{H}$ is the channel matrix in delay-time domain with the elements $[\bm{H}]_{i,j}=h[i-j,i]$ for $i,j=0,\ldots,MN+L_{\rm{cp}}-1$. 

Considering the demodulator structure in Fig.~\ref{fig:OTFS_Rx1}, the received delay-Doppler domain data symbols can be obtained as
 \bea \label{eqn:dd_rec1}
\widetilde{\bm{d}} 
&\!\!=\!\!&\bm{H}^{ \mathtt{OTFS}}_{\rm{DD}}\dd+\boldsymbol{\widetilde{\eta}}_{ \mathtt{OTFS}},
\eea
where $\boldsymbol{\widetilde{\eta}}_{ \mathtt{OTFS}} = \bOmega^{\rm{H}} (\bm{I}_{N} \otimes \bm{F}^{\rm{H}}_{M}) \bPsi^{\rm{H}} \bm{F}_{MN} 
\boldsymbol{\eta}$ is the noise vector at the OTFS demodulator output and $\bm{H}^{ \mathtt{OTFS}}_{\rm{DD}}=\bOmega^{\rm{H}} (\bm{I}_{N} \otimes \bm{F}^{\rm{H}}_{M}) \bPsi^{\rm{H}} \bm{F}_{MN} \BH \bm{F}^{\rm{H}}_{MN} \bPsi (\bm{I}_{N} \otimes \bm{F}_{M}) \bOmega$ is the equivalent delay-Doppler domain channel for OTFS.

As it was mentioned in the previous section, the only difference between OTFS and SC-IFDMA is in the absence of the phase factors at the input of the DFT precoders and output of the IDFT post-processors of the transmitter and receiver, respectively. Accordingly, the input-output relationship for data transmission using SC-IFDMA can be represented as
 \be \label{eqn:dd_rec_SC-FDMA}
\widetilde{\bm{d}}_{ \mathtt{SC-IFDMA}} =\bm{H}^{ \mathtt{SC-IFDMA}}_{\rm{DD}}\dd+\boldsymbol{\widetilde{\eta}}_{ \mathtt{SC-IFDMA}},
\ee
where $\boldsymbol{\widetilde{\eta}}_{ \mathtt{SC-IFDMA}} = (\bm{I}_{N} \otimes \bm{F}^{\rm{H}}_{M}) \bPsi^{\rm{H}} \bm{F}_{MN} 
\boldsymbol{\eta}$ and $\bm{H}^{ \mathtt{SC-IFDMA}}_{\rm{DD}}=(\bm{I}_{N} \otimes \bm{F}^{\rm{H}}_{M}) \bPsi^{\rm{H}} \bm{F}_{MN} \BH \bm{F}^{\rm{H}}_{MN} \bPsi (\bm{I}_{N} \otimes \bm{F}_{M})$ is the equivalent channel for SC-IFDMA. From (\ref{eqn:dd_rec_SC-FDMA}) and (\ref{eqn:dd_rec1}), it can be easily understood that
\be \label{eqn:ch_rel}
\bm{H}^{ \mathtt{SC-IFDMA}}_{\rm{DD}}= \bOmega \bm{H}^{\mathtt{OTFS}}_{\rm{DD}} \bOmega^{\rm{H}}.
\ee
Therefore, the elements of OTFS and SC-IFDMA channel matrices have the same amplitude and they are only different in known phases. On this basis, we conclude that the end-to-end channel response for SC-IFDMA is also a dela-Doppler representation of the channel. Thus, similar to OTFS \cite{OTFS}, SC-IFDMA is also capable of converting the LTV channel to a 2D time-invariant one in the delay-Doppler domain. Additionally, since $\bOmega$ is a unitary matrix, the same detection performance for both systems is expected when the same pilot signals, and detection techniques are deployed for both systems. To confirm this, in the following section, we compare the detection performance of both systems by simulations.

\section{Detection}
\label{sec:Detection}
In this section, we numerically compare the detection performance of SC-IFDMA and OTFS. We consider synchronization, channel estimation, and equalization in the detection process. Additionally, we propose a fine-tuning technique for time synchronization that provides highly accurate timing estimates for both OTFS and SC-IFDMA. For synchronization and channel estimation of both waveforms, the widely used impulse pilot at a given delay-Doppler bin $(m_{\rm p}, n_{\rm p})$ with the power $\rho_{\rm p}$, surrounded by zero guards is considered \cite{Raviteja2019}.

It is worth noting that preamble signals or frequency domain pilots have been deployed for synchronization and channel estimation in SC-FDMA literature so far \cite{Falconer2002,Liu2015,Kiayani2016}. Hence, utilization of embedded pilots among data symbols before DFT precoding for SC-FDMA is proposed for the first time in this paper, thanks to the derivations in Section~\ref{sec:Channel_Effect}.

Let us consider the received SC-IFDMA/OTFS signal in presence of synchronization errors as
\be \label{eqn:usyn_rec}
r[\kappa]=e^{\frac{j2 \pi \varepsilon k}{MN}} \sum_{\ell=0}^{L_{\rm{ch}}-1} h[\ell,\kappa] x[\kappa-\ell-\theta]+\eta[\kappa], 
\ee
where $\theta=\theta_{\rm d}+M\theta_{\rm t}$ and $\varepsilon$ are the normalized TO and CFO values to the delay and Doppler spacings with $\theta_{\rm d}$ and $\theta_{\rm t}$ denoting the TO in delay and time dimensions, respectively.

\subsection{Synchronization}\label{subsec:synch}

\begin{figure*}[!t]
\begin{multicols}{3}
\centering 
    \includegraphics[scale=0.21]{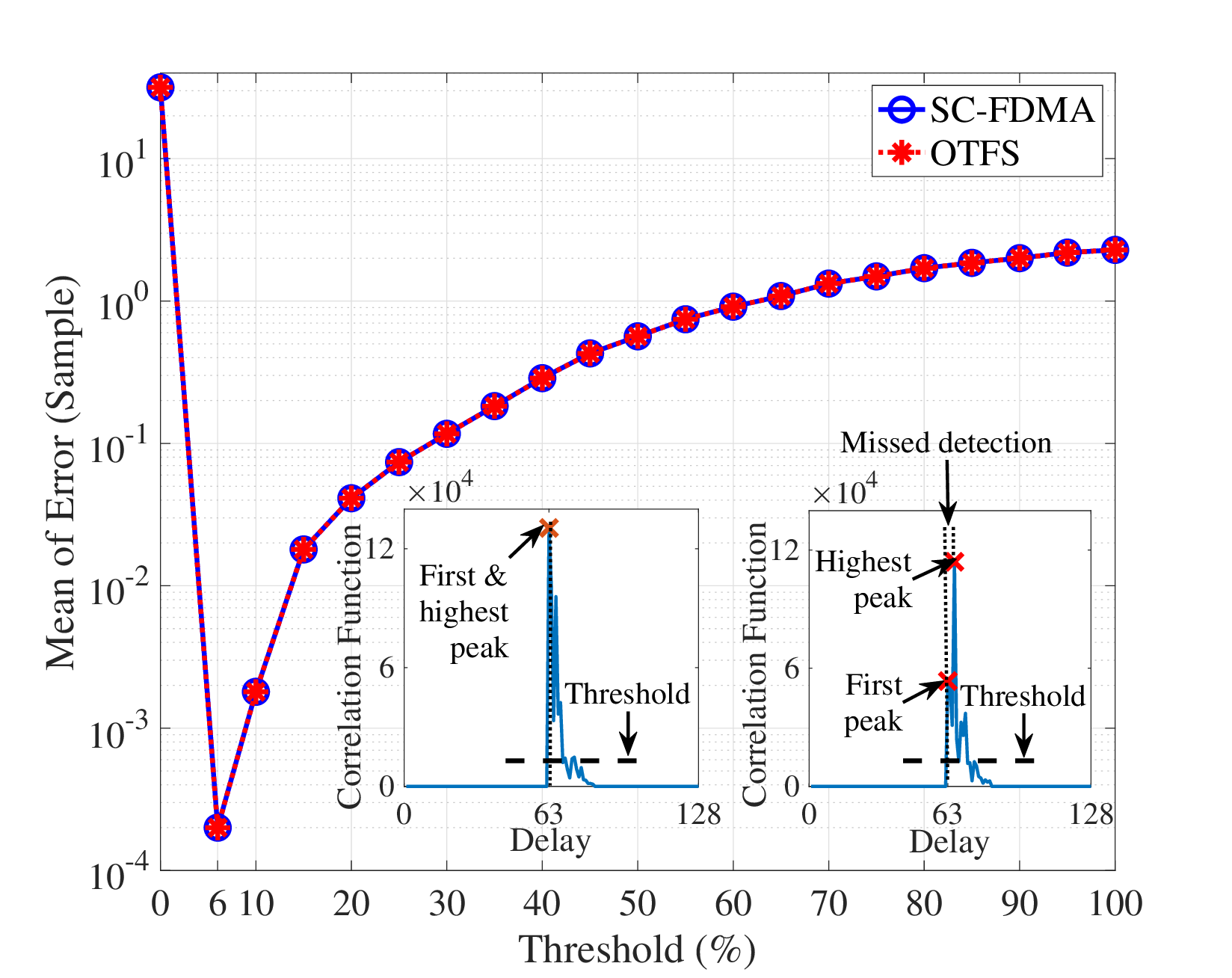}\par \vspace{-0.3cm} \caption{Effect of threshold selection on the accuracy of TO estimation in the delay dimension.} \label{fig:peak}
  \includegraphics[scale=0.21]{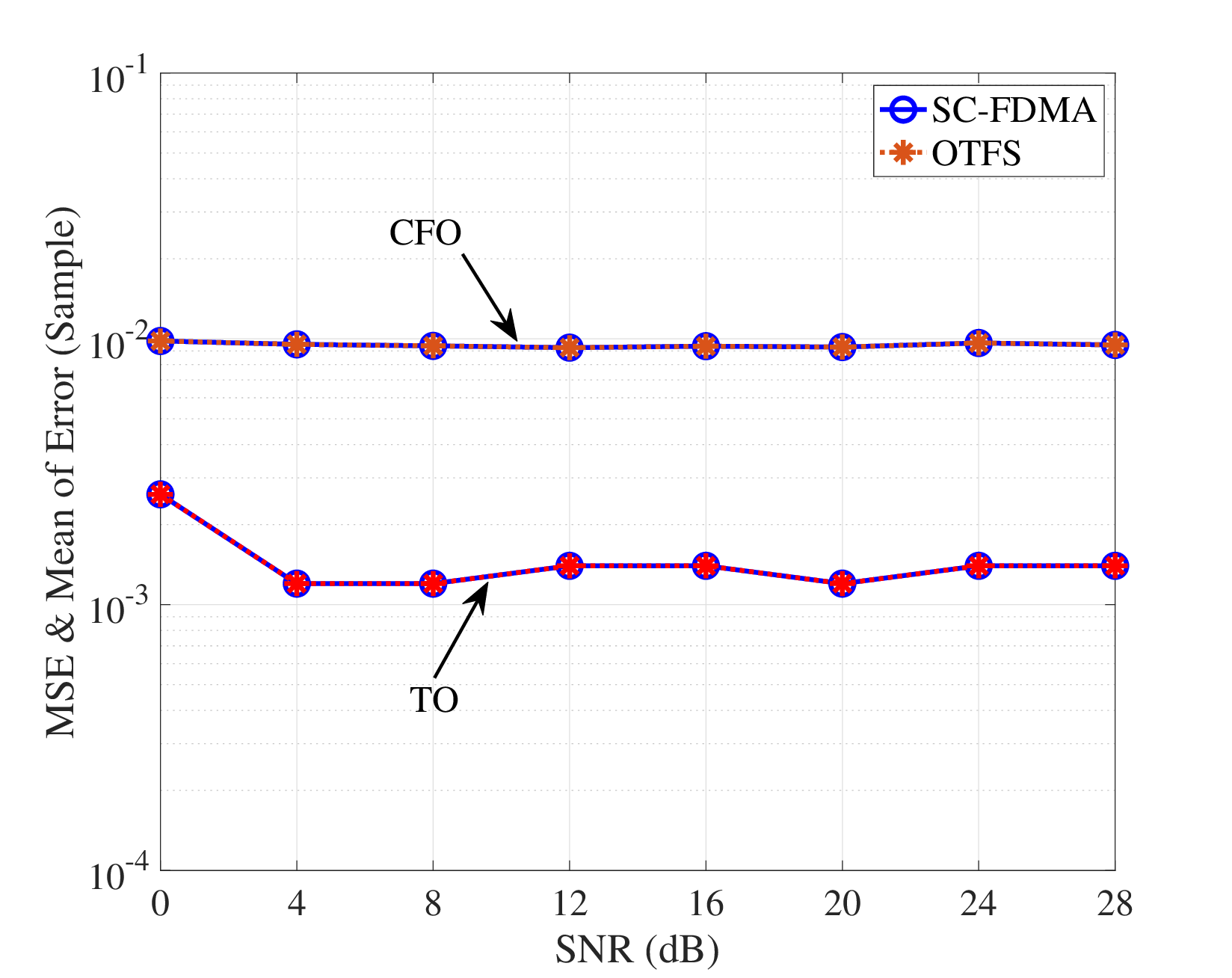}\par \vspace{-0.3cm} \caption{MSE and mean error for CFO and TO estimation in SC-FDMA and OTFS, respectively.} \label{fig:sync}
    \includegraphics[scale=0.21]{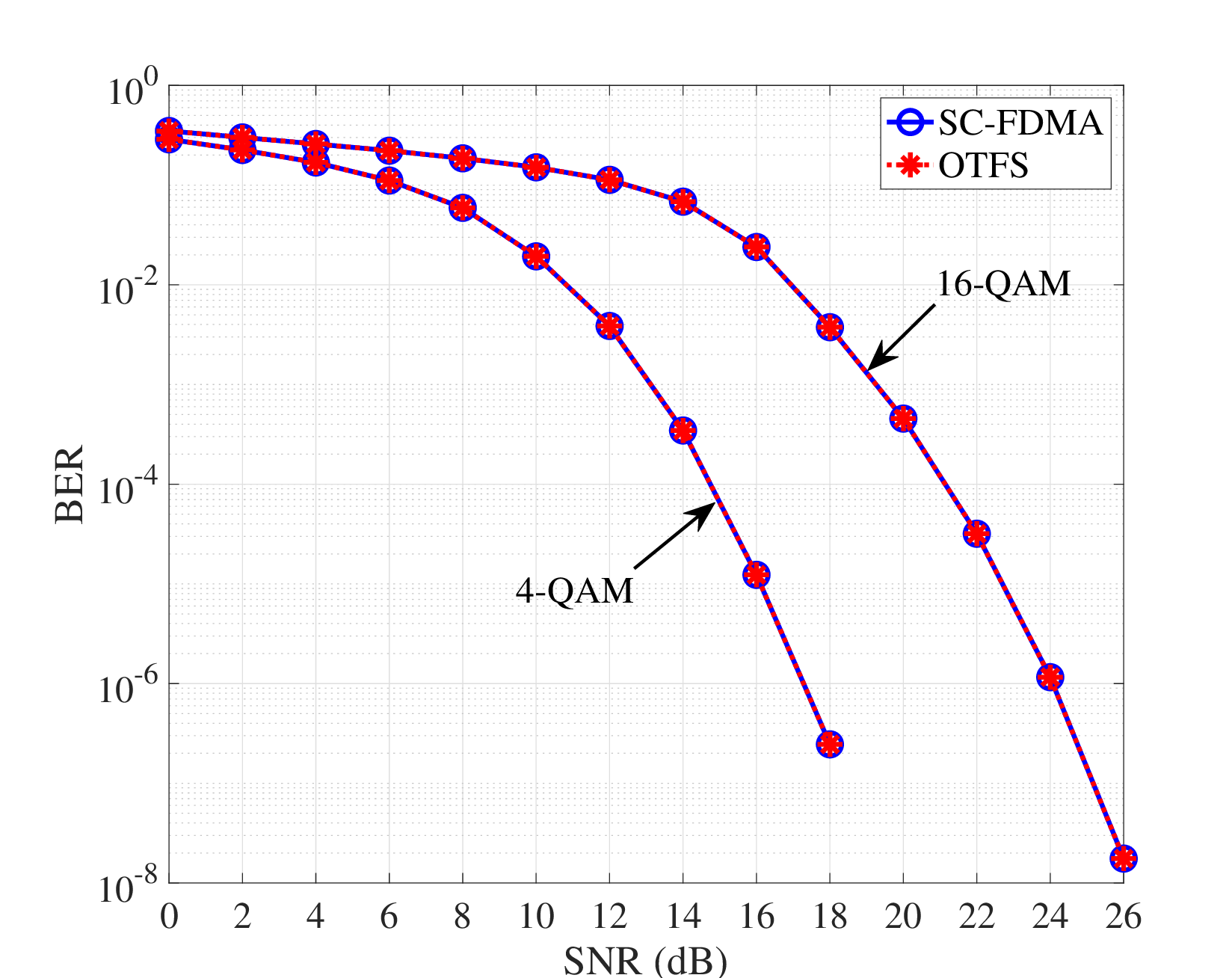}\par \vspace{-0.3cm} \caption{BER performance of impulse pilot channel estimation for SC-FDMA and OTFS.} \label{fig:ch}
\end{multicols}
\vspace{-0.7 cm}
\end{figure*}
To estimate the TO, we need to search for the pilot sequence on a row of the delay-time grid. Thus, we convert the received signal from serial to parallel, with blocks of $M$ samples in each parallel stream representing the samples on the columns of the grid. Consequently, we rearrange the received signal as $r[m,l]=r[Ml+m]$ with the delay and time indices $m$ and $l$, respectively. Similar to \cite{Bayat2022}, for a given row $m$ on the delay-time grid, we consider a sliding window with length $N$ that searches for the pilot sequence with $N$ identical samples. 
Thus, we use the same timing metric as in \cite{Bayat2022}, i.e., $P[m,l]=\sum_{q=0}^{N-2} r^*[m,l+q] r[m,l+q+1]$.
Based on the results of \cite{Bayat2022}, considering the CP and pilot position in delay, ${\theta}_{\rm d}$ can be estimated by finding the peak of the timing metric $P_{\rm d}[m]\!\!=\!\!\sum_{l=0}^{N-1} \!\!P[m,l]$ as
\be \label{eqn:TO_d1} 
\hat{\theta}_{\rm d}={\arg} \max_m \big\{ |P_{{\rm d}}[m]| \big\}-m_{\rm{p}}-L_{\rm{cp}}.
\ee
However, the multipath effect of the channel introduces a bias in the TO estimate. This bias can be partially corrected with the knowledge of the first-order moment of the channel \cite{Bayat2023}. Nevertheless, the fractional value of this moment may lead to inaccurate TO estimation. Furthermore, the prior knowledge of this moment may not be available. Hence, we propose a fine-tuning technique after course TO estimation in (\ref{eqn:TO_d1}).

As it is explained in \cite{Bayat2022} and \cite{Bayat2023}, the peak of the correlation function in (\ref{eqn:TO_d1}) is dominated by the maximum tap of the channel. Consequently, when the first tap is not the largest, this leads to estimation error. To tackle this issue, we propose to refine the course TO estimate in (\ref{eqn:TO_d1}) by identification of the first peak of the correlation function instead of its maximum peak. To this end, after finding the maximum peak, we set a threshold, $0<\mathcal{T}_{\rm{s}}\leq1$, to determine the values constituting the group of peaks of the correlation function within a percentage of the maximum peak value. Using this threshold, the sample indices for the peaks of $P_{\rm d}[m]$ are stored in a set
\be \label{eqn:TO_d} 
\boldsymbol{\widehat{\Theta}} \!=\! \Big\{ m  \Big| \big|P_{{\rm d}}[m] \big| \!\geq\! \big( \mathcal{T}_{\rm{s}} \times \max \big\{ |P_{{\rm d}}[m]| \big\} \big) \Big\}.
\ee
Subsequently, the first peak which provides more accurate TO estimate is obtained as 
\be \label{eqn:TO_d2} 
\hat{\theta}_{\rm d}^{\mathtt{fine}}= \min \{ \boldsymbol{\hat{\Theta}} \}-m_{\rm{p}} - L_{\rm{cp}}.
\ee
Finally, $\theta_{\rm t}$ and $\varepsilon$ can be estimated using the method in \cite{Bayat2022}.

In the following, we show the efficacy of our proposed fine TO estimation technique for both OTFS and SC-IFDMA using simulations. We consider $M=128$ delay bins, $N=32$ Doppler bins at the carrier frequency of $f_{\rm{c}}\!=\!5.9$~GHz, and the bandwidth ${\rm BW}=7.68$~MHz. A CP  longer than the channel delay spread is appended at the beginning of each block. The extended vehicular~A (EVA) channel model \cite{3gpp} with the relative velocity of $v=500$~km/h between the transmitter and receiver is considered.

In Fig.~\ref{fig:peak}, we analyze the mean of error for our proposed fine TO estimation technique as a function of $\mathcal{T}_{\rm{s}}$. It is worth noting that $\mathcal{T}_{\rm{s}}=1$ corresponds to the technique in \cite{Bayat2022} with one to two sample errors on average. As shown, decreasing the threshold leads to orders of magnitude more accurate TO estimates compared to the course estimation technique in \cite{Bayat2022}. This is due to the fact that a lower threshold increases the chance of finding the first peak of $P_{\rm d}[m]$. However, setting a very small threshold leads to performance degradation which is due to the noise effect. Fig.~\ref{fig:peak} also depicts the correlation function for two realizations of the channel. In Fig~\ref{fig:sync}, we compare the CFO and TO estimation performance for $16$-QAM in terms of mean square error (MSE) and mean of sample errors, respectively, versus signal to noise ratio (SNR) where both waveforms perform similarly. As shown in Figs.~\ref{fig:peak} and \ref{fig:sync}, both OTFS and SC-IFDMA exhibit the same synchronization performance.

\subsection{Channel Estimation and Equalization}
Considering (\ref{eqn:ch_rel}), the equivalent delay-Doppler domain channel for SC-IFDMA only differs from that of OTFS in phase and not the amplitude. 
Moreover, this phase difference does not affect the magnitude of the received pilot. Therefore, channel estimation techniques used for OTFS, such as the one in \cite{Raviteja2019}, can be applied to SC-IFDMA. 
In this approach, the received pilot copies in the delay-Doppler domain with magnitudes above a positive detection threshold are retained and divided by the transmitted pilot to obtain the delay-Doppler channel estimate for both SC-FDMA and OTFS.

Based on the results and insights provided in Sections~\ref{sec:OTFS_SC-FDMA} and \ref{sec:Channel_Effect}, SC-FDMA is in fact a delay-Doppler multiplexing technique. Hence, we propose to perform channel equalization after the IDFT post-processing units by using the same equalization techniques as the ones that are used for OTFS. This is a different approach to the existing channel equalization techniques in SC-FDMA literature \cite{Falconer2002,Kiayani2016,Goodman2006}. With this approach, the same gains as those for OTFS are expected to be achieved by SC-FDMA. To evaluate this, in Fig.~\ref{fig:ch}, we compare the BER performance of OTFS with SC-IFDMA when the channel estimation and equalization techniques in \cite{Raviteja2019} and \cite{LSMR2021} that were originally proposed for OTFS are applied to both systems. As shown, the BER performance of SC-IFDMA perfectly matches with that of OTFS using the same parameters as the ones in Section~\ref{subsec:synch}. 

\section{Mulitiple Access}\label{sec:Multi_User_v0}
Multiple access is an important aspect of modern communication systems. While the literature on delay-Doppler multiplexing in single-user scenarios is quite rich, \cite{Wei2021}, there are only a limited number of works that investigate its multiple access aspects \cite{SaifKhan2019, Chockalingam2019, SaifKhan2022, Das2023}. In contrast, there is a rich literature behind SC-FDMA and in fact finding the relationship between OTFS and SC-FDMA, in this paper, opens interesting avenues for both waveforms. Therefore, in the following, we derive an equivalent multiuser channel in the uplink direction that incorporates the channel effects of all the users. Without loss of generality, based on the results of Section~\ref{sec:Channel_Effect}, we represent the channel matrix for a given user when SC-FDMA or OTFS is deployed as $\bH_{\rm{DD}}^q$.

\begin{figure}
\psfrag{T}{ \hspace{-1.3mm}$T$}
\psfrag{A}{ \hspace{-1.3mm}$NT$}
\psfrag{t}{ \hspace{-1.3mm}$\tau$}
\psfrag{n}{ \hspace{-1.8mm}$\nu$}
\psfrag{f}{ \hspace{-1.3mm}$f$}
\psfrag{C}{ \hspace{-1.3mm}$t$}
\psfrag{F}{ \hspace{-2mm}$\Delta f$}
\psfrag{B}{ \hspace{-1.3mm}$M\Delta f$}
\psfrag{D}{ \Large\hspace{-1.3mm}$\mathcal{U}_0$}
\psfrag{E}{ \Large\hspace{-1.3mm}$\mathcal{U}_1$}
\psfrag{G}{ \Large\hspace{-1.3mm}$\mathcal{U}_2$}
  \centering 
  {\includegraphics[scale=0.3]{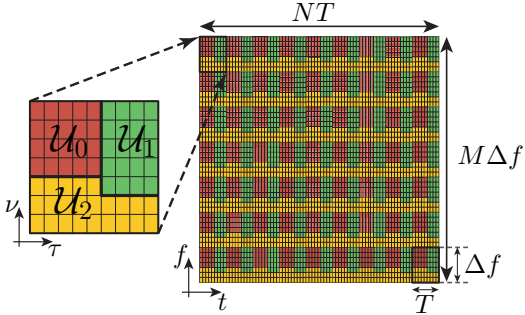}}
    \vspace{-4 mm}
  \caption{Delay-Doppler vs. time-frequency domain multiple access.}
    \vspace{-5 mm}
  \label{fig:mu}
\end{figure}

We consider $Q$ users sharing $M$ delay bins and $N$ Doppler bins with $\mathbb{U}_\tau^q$ and $\mathbb{U}_\nu^q$ being the sets of $M_q$ delay bins and $N_q$ Doppler bins allocated to the users $q\!=\!0,\ldots,Q\!-\!1$, where no single delay-Doppler resource is used by more than one user, i.e., $\mathbb{U}_{\tau\!/\!\nu}^i\cap\mathbb{U}_{\tau\!/\!\nu}^j=\emptyset$ for $i\neq j$.
We define the $M_q\times N_q$ data matrices of the users $q=0,\ldots,Q-1$ as $\BD^q$ where $M=\sum_{q=0}^{Q-1}M_q$ and $N=\sum_{q=0}^{Q-1}N_q$. The delay and Doppler resource allocation matrices $\bGamma_\tau^q$ and $\bGamma_\nu^q$ are formed by the columns of $\I_{M}$ with the indices that belong to the set $\mathbb{U}_\tau^q$ and the rows of  $\I_{N}$ with the indices from the set $\mathbb{U}_\nu^q$, respectively. Using these matrices, the data symbols of each user $\BU_q$ are mapped to their corresponding delay-Doppler bins as $\bD_q=\bGamma_\tau^q \BD^q \bGamma_\nu^q$. Here, we consider generalized resource allocation and thus, the delay-Doppler resources can be allocated to the users without any particular pattern. Based on the same principles as in Fig.~\ref{fig:DD_TF}, the delay-Doppler and time-frequency resources that are occupied by different users (known as tiles) are shown in Fig.~\ref{fig:mu}.

After vectorizing $\bD_q$, i.e., ${\rm vec}\{\bD_q\}=\bGamma^q\y_q$ where $\bGamma^q=(\bGamma^{q}_{\nu})^{\rm{T}} \otimes \bGamma_\tau^q$ and $\y_q={\rm vec}\{\BD_q\}$. By replacing $\BH$ with $\BH^q$, i.e., the channel matrix of user $q$, in (\ref{eqn:dd_rec1}) or (\ref{eqn:dd_rec_SC-FDMA}), the received signal from this user at the base station (BS) is ${\bm{\widetilde{y}}}_{q} = \bm{H}_{\rm{DD}}^q \y_q$ where $\bm{H}_{\rm{DD}}^q$ is either the OTFS or SC-IFDMA channel matrix of user $q$. Hence, the combined received signal from all the users at the BS can be represented as 
\be \label{eqn:dd_rec}
\vspace{-1 mm}
{\bm{\widetilde{d}}} = \sum_{q=0}^{Q-1} \bm{H}_{\rm{DD}}^q \y_q + \boldsymbol{\widetilde{\eta}} = \bm{H}_{\rm{DD}} \bm{d} + \boldsymbol{\widetilde{\eta}},
\vspace{-1 mm}
\ee
in which $\bm{H}_{\rm{DD}}=\sum_{q=0}^{Q-1} \bm{H}_{\rm{DD}}^q \bm{\Gamma}^q$ is the compound channel matrix that includes the channel responses of all the users, $\dd\!\!=\!\! \sum_{q=0}^{Q-1}\y_q$ and $\bm{\widetilde{d}} \!\!=\!\! \sum_{q=0}^{Q-1} {\bm{\widetilde{y}}}_{q}$. Using the result in (\ref{eqn:dd_rec}), the existing OTFS detectors, e.g., \cite{LSMR2021}, can be utilized to estimate the transmitted signals of all the users for both OTFS and SC-IFDMA. As it was shown in the previous section, the same performance for both waveforms can be achieved. Further investigation of multiple access OTFS/SC-IFDMA with our perspective in this paper is left as future work.

\section{Conclusion}\label{sec:Conclusion}
In this paper, we compared OTFS and SC-FDMA by mathematical and numerical analysis. This analysis led us to discover that SC-FDMA is a delay-Doppler multiplexing technique. We also showed that SC-FDMA achieves the same performance gains as OTFS in LTV channels. This is a promising result as SC-FDMA is a part of the current wireless standards. This study also led to deep insights into the resource allocation aspects of OTFS. We also developed an accurate TO estimation applicable to both waveforms. Finally, we derived input-output relationships for the uplink communication channel with generalized resource allocation.

\section*{Acknowledgement}\label{sec:Acknowledgement}
This publication has emanated from research conducted with the financial support of Science Foundation Ireland under Grant numbers SFI/19/FFP/7005(T) and SFI/21/US/3757.

\bibliographystyle{IEEEtran} 

\end{document}